\documentstyle[aps,prl] {revtex}
\begin{document}
\draft
\twocolumn[\hsize\textwidth\columnwidth\hsize\csname %
@twocolumnfalse\endcsname
\title{Li$_2$VO(Si,Ge)O$_4$, a prototype of a two-dimensional frustrated 
quantum Heisenberg antiferromagnet}


\author{
R. Melzi $^1$, P. Carretta$^1$, A. Lascialfari$^1$, M. Mambrini$^2$, M. Troyer$^3$
P. Millet$^4$ and F. Mila$^2$}
\address{
$^1$ Dipartimento di Fisica "A. Volta", Unit\'a INFM  di Pavia,
27100 Pavia, Italy}
\address{
$^2$ Laboratoire de Physique Quantique, Universit\'e Paul Sabatier, 31062 
Toulouse Cedex, France
} 
\address{
$^3$ Institute of Theoretical Physics, ETH H\"onggerberg, CH-8093 Z\"urich, Switzerland
} 
\address{
$^4$ Centre d' Elaboration des Mat\'eriaux et d' Etudes Structurales, CNRS, 
31055 Toulouse Cedex, 
France
} 
\date{\today}
\maketitle
\widetext
\begin{abstract}


NMR and magnetization measurements in Li$_2$VOSiO$_4$ and Li$_2$VOGeO$_4$ are reported.
The analysis of the susceptibility shows that both
compounds are two-dimensional $S=1/2$ Heisenberg antiferromagnets
on a square lattice with a sizeable frustration induced by 
the competition between the superexchange couplings $J_1$ along the sides of 
the square 
and $J_2$ along the diagonal.
Li$_2$VOSiO$_4$ undergoes a low-temperature phase transition to a collinear order,
as theoretically predicted for $J_2/J_1 > 0.5$. Just above the magnetic transition 
the degeneracy between the two collinear ground states is lifted by the onset
of a structural distortion. 















\end{abstract}
\newpage
%
]
\narrowtext


In recent years one
has witnessed an extensive investigation of quantum phase transition in low-dimensional
$S=1/2$ Heisenberg antiferromagnets (QHAF) as a function of doping, magnetic field and 
disorder\cite{QHAF}. For example, two-dimensional QHAF (2DQHAF) 
have been widely 
studied  to show the occurrence of
a phase transition from the renormalized
classical to the quantum disordered regime 
upon charge doping \cite{HTCSC}.  Another possibility to drive quantum 
phase transitions in a 2DQHAF
is to induce a sizeable
frustration. In particular, for a square lattice with an exchange coupling along the
diagonal $J_2$ about half of the one along the sides 
of the square $J_1$ (see Fig. 1a),
a crossover to a quantum disordered phase with a finite gap between 
the singlet ground state
and the first excited state is expected \cite{Chandra1,Schulz,Sorella}. 
For $J_2/J_1\ll 0.5$ a N\'eel order is envisaged, while
for $J_2/J_1\gg 0.5$ a collinear order should develop. 
The collinear order (see Fig. 1a), which
 can be considered as formed by two interpenetrating N\'eel 
sublattices with staggered magnetization ${\bf n_1}$ and  ${\bf n_2}$, is characterized
by an Ising order parameter 
$\sigma= {\bf n_1}. {\bf n_2} =\pm 1$  \cite{Chandra2}. The two 
values of $\sigma$ correspond to the two  collinear  configurations,
one with spins ferromagnetically
aligned along the $x$ axis, with a magnetic wave-vector 
${\bf Q}=(0, \pi/a)$,
the other with spins ferromagnetically aligned along the $y$ axis
$({\bf Q}=(\pi/a,0))$. At a certain temperature an Ising phase transition occurs and
the system choses among the $x$ or $y$ collinear configurations.
The precise boundaries of the $J_2/J_1$
phase diagram for a frustrated 2DQHAF are unknown and could be modified 
by the presence of a finite third
neighbour coupling \cite{Chandra2}. These theoretical predictions
have not found  an experimental
support so far, 
mainly due to the absence of a system which can be regarded as a prototype
of a frustrated 2DQHAF. 


In this letter we present NMR and magnetization measurements that
prove that the isostructural compounds Li$_2$VOSiO$_4$ (LSVO for short) 
and Li$_2$VOGeO$_4$ (LGVO) \cite{Millet}, formed by layers of
V$^{4+}$ ($S=1/2$) ions on a square lattice (see Fig. 1b), are prototypes
of frustrated 2DQHAF with significant coupling between both first ($J_1$) 
and second ($J_2$) neighours.
Moreover we show that LSVO undergoes a phase transition to a low temperature 
collinear order,
as expected for $J_2/J_1> 0.5$. The phase transition is triggered by a lattice
distortion which lifts the degeneracy between the two possible collinear ground 
states and
could belong to the Ising universality class. 


$^{29}$Si NMR and magnetization measurements have been performed on 
powder samples
while $^7$Li NMR measurements, thanks to $^7$Li sensitivity, 
have been carried out also on a $\sim 1\times 1\times 0.2$ mm$^3$ LSVO single crystal.  
NMR spectra and nuclear spin-lattice relaxation rate $1/T_1$ have been
measured by using standard pulse sequences.
The field-cooled magnetization $M$ was measured 
with a commercial Quantum Design MPMS-XL7 SQUID magnetometer.


The structure of $V^{4+}$ layers \cite{Millet}
suggests that both the couplings between 
first and second neighbours can be significant. It is however difficult
a priori to decide which one should dominate: first neighbours are 
connected by two superexchange channels, but they are located in pyramids looking 
in opposite directions and are not exactly in the same plane, whereas second neighbours are
connected by only one channel, but they are located in pyramids looking 
in the same directions and are in the same plane. It would thus be highly desirable 
to extract information on the relative value of these
exchange integrals from the susceptibility ($\chi =M/H $) (see Fig. 2). Although
 the temperature dependence of the susceptibility of the 
$J_1-J_2$ model is not known accurately as a function of $J_1$ and $J_2$, it turned out to be 
possible to obtain useful information from the following considerations. If the system was not 
frustrated, i.e. if $J_1\gg J_2$ or $J_2\gg J_1$, the susceptibility would be that
of a regular Heisenberg AF on the square lattice with coupling $J$. In that case, 
Quantum Monte Carlo simulations can be used to determine the temperature dependence 
of the susceptibility\cite{Troyer}, and the maximum occurs at 
$T_{\rm max}\simeq 0.935 J$. Since in that case the Curie-Weiss 
temperature
$\Theta= J$ one has a ratio $T_{\rm max}/\Theta\simeq 0.935$.
Now, while $T_{\rm max}$ is known very accurately from our measurements, the precise
determination of $\Theta$ is more problematic. A simple fit with 
$\chi (T)= \chi_{VV} + C/(T+\Theta)$, with $\chi_{VV}$ Van-Vleck susceptibility, 
is not good enough because there is an important 
dependence of the results on the lowest temperature used in the fit. To overcome this 
problem, we have performed a fit of the high
temperature part of the susceptibility up to third order. The coefficients are 
then consistent with a $J_1-J_2$ model only in a small window for the lowest 
temperature. Within this window, $\Theta$ depends very weakly on 
the lowest temperature and a precise estimate
of the Curie-Weiss temperature can be achieved. 
The results are $\Theta \simeq 7.4 K$ for LSVO and 
$\Theta\simeq 5.2 K$ for LGVO \cite{Notachi}. 
Accordingly, the ratios $T_{\rm max}/\Theta$ are equal
to 0.72 and 0.67, respectively. In both cases, this ratio is significantly lower than
the value 0.935, supporting the presence of a 
sizeable frustration. Besides, with a smaller ratio, LGVO is
expected to be closer to the fully frustrated point $J_2/J_1=1/2$ than LSVO, in 
qualitative agreement with the fact that no phase transition was found in that system
down to 1.9 K (see below). What we cannot say however on the basis
of this analysis is which of the couplings $J_1$ and $J_2$ is larger.
  
  
To be more quantitative, we need to know the ratio $T_{\rm max}/\Theta$ for 
the $J_1-J_2$ model as a function of $J_2/J_1$. It turned out to be impossible
to get accurate estimates in the strongly frustrated region $J_2/J_1\simeq 1/2$, but 
exact diagonalizations with 3 sizes available (4, 8 and 16 sites) give a reliable 
estimate for $J_2/J_1<.4$, while Quantum Monte Carlo simulations, which suffer from 
the minus sign problem, provide useful information down to $J_2/J_1\simeq 2$ on 
the other side, where only two sizes can be used for exact diagonalizations due to the
type of order
(8 and 16 sites). The experimental ratio $T_{\rm max}/\Theta =.72$ for LSVO 
then implies
that $J_2/J_1$ is approximately equal to .1 or to 3.5, while for LGVO, 
$T_{\rm max}/\Theta =.67$ implies that $J_2/J_1$ is either close to .25 or to 2.5.






A significant difference between the
two compounds is discernable if one reports the derivative $d\chi/dT$ vs. T 
(see the inset
to Fig. 2). One observes that around $T_c\simeq 2.83$ K a peak is present for LSVO, 
while
no anomaly in $d\chi/dT$ is detected for LGVO, down to 1.9 K. 
The peak occurs at the same temperature
where a peak in $^7$Li NMR $1/T_1$ is observed (see Fig. 3), 
signaling a phase transition to a magnetically ordered state.
Remarkably, $T_c$ was found independent on the magnetic field intensity, within $\pm 0.15$ K (i.e. $\pm
5$\% ), up to $H=7$ Tesla.


In LSVO, for $H= 1.8$ Tesla, one observes that 
$^7$Li $1/T_1$ is constant between $3.5$ and $293$ K.
In the high temperature limit ($T\gg \Theta$), 
by resorting to the usual  Gaussian form for the spin correlation
function one has \cite{Moriya}
\begin{equation}
(1/T_1)_{\infty}={\gamma^2\over 2}{S(S+1)\over 3}{\sqrt{2\pi}\over \omega_E}\times
\sum_{k,i,j}  (A_{ij}^k)^2
\end{equation}
with $A_{ij}$ ($i= x,y,z$, $j=x,y$) 
the components of the hyperfine tensor due to the $k^{th}$ V$^{4+}$, 
$\gamma$ the gyromagnetic ratio and
$\omega_E={Jk_B}\sqrt{2 z S(S+1)/3}/\hbar$.
$z=8$ is the number of nearest neighbour spins of a V$^{4+}$ coupled via an effective
superexchange coupling $J$ related to $J_1$ and $J_2$. 
$^7$Li hyperfine coupling constants in LSVO have been estimated by reporting 
the temperature dependence of the paramagnetic shift, both for the single crystal
and for the powders, as a function of the susceptibility. 
It turned out that $^7$Li nuclei are coupled to V$^{4+}$ ions
both via a dipolar and a transferred hyperfine
coupling $A_T= 1700$ Gauss \cite{Melzi}, which is attributed to the two V$^{4+}$
nearest neighbours. The dipolar term is close to the one
estimated on the basis of lattice sums.
From Eq. 1, by using the
high temperature value of $1/T_1$, one finds $J= 6.4$ K, a value consistent with 
the Curie-Weiss temperature 
derived from the analysis of the susceptibility. 


For $T <\Theta$, V$^{4+}$ spins are strongly correlated and the 
behaviour of $1/T_1$ depends on which regime LSVO is: 
quantum critical,
renormalized classical or quantum disordered \cite{Rep}. However, one has to notice that
the temperature dependence of $1/T_1$ can be strongly influenced by the $q$-dependent
hyperfine form factor \cite{CFTD}.
Hence, in order to make significant statements on the correlated spin dynamics of 
this frustrated 2DQHAF we have first calculated
the hyperfine form factor, which was found only weakly $q-$dependent in view of the sizeable
transferred coupling. Therefore, the temperature dependence of $1/T_1$ is fully determined by 
the one of the correlation length and the fact that $1/T_1$ is constant
down to $\simeq 3$ K suggests that LSVO, for $H=1.8$ Tesla, is in the quantum critical regime
\cite{Rep}.


Below $T_c$ one observes an activated decrease of $1/T_1$ which is typical of a
magnetically ordered system with a gap in the spin wave spectrum \cite{Pincus}. An estimate of
the gap amplitude can be done using a simple Arrhenius fit, which yields $\Delta= 18$ K
(see the inset to Fig. 3).
Although the estimate of $\Delta$ might not be very accurate in view of a certain proximity to
the phase transition it should be noticed that 
the estimated value is considerably larger
than what one would expect for a magnetic system with a coupling constant of
a few degrees kelvin, as deduced from susceptibility and $1/T_1$ measurements for 
$T\gg \Theta$. This fact could suggest a modification of the superexchange coupling
at low temperatures, possibly involving a lattice distortion. The occurrence of a
lattice distortion is, in fact, corroborated by the modifications in $^{29}$Si NMR powder
spectra below $\simeq 3.4$ K (see Fig. 4a). One observes, just above $T_c$
the appearence of a shifted narrow peak in $^{29}$Si  NMR powder spectrum. 
On decreasing temperature
the  low-frequency peak progressively disappears while the intensity of the high frequency one
increases. Two aspects should be remarked: $1)$ $^{29}$Si NMR line does not broaden
below $T_c$ indicating that the local field at $^{29}$Si nuclei is zero; $2)$ the modification
in the shift has to be associated with a modification in the chemical shift or hyperfine 
coupling, suggesting
the occurrence of a structural distortion. The absence of a magnetic field at $^{29}$Si site
can be accounted for either by an AF order with the spins parallel to the $c$ axis
or by a collinear order with  V$^{4+}$ spins along the sides of the square lattice.
In order to exclude the presence of an AF order we studied the angular dependence of the magnetic field
at $^7$Li nuclei below $T_c$ \cite{Melzi}. By considering the same hyperfine coupling tensor
determined above $T_c$  \cite{NotaLi} 
we found that for the AF order the magnetic field intensity should
always increase on turning the magnetic field from parallel to the $c$-axis to parallel
to the $ab$ plane, at variance with the experimental findings. Moreover, the fact that
the spins are parallel to the $ab$ plane below $T_c$ is  supported by a recent EPR
analysis of the $g$ tensor \cite{Stepanov}, showing that there is a larger in-plane
magnetic anisotropy.
Therefore, we conclude that the magnetic order is collinear with the spins in the $ab$ plane.
{\it This is the first evidence of a collinear order in a frustrated 2DQHAF, whose
existence has been theoretically put forward long ago by Chandra and coworkers 
\cite{Chandra1,Chandra2}}.
It should be observed that only one of the two possble collinear orders, with $\sigma=\pm 1$,
is compatible with zero magnetic field at $^{29}$Si, the one with the spins parallel
to the staggered modulation, i.e. for spins along the $x$-axis the one with magnetic
vector ${\bf Q}=(\pi/a, 0)$. The fact that in LSVO always one type of collinear order
develops indicates that the four-fold symmetry of the square lattice was broken, possibly
by the lattice distortion occurring just above $T_c$.



Further information on the collinear phase in LSVO can be derived from the temperature dependence
of $^{7}$Li NMR spectra below $T_c$ (see Fig. 4c). 
The temperature dependence of the order parameter, proportional to
V$^{4+}$ average magnetic moment, 
is obtained from the splitting of the satellites of $^7$Li NMR spectrum. 
The central peak and the two satellites do not correspond 
to the $1/2 \to -1/2$ and $\pm 3/2\to \pm 1/2$ transitions, respectively, but 
correspond to Li sites where the local field is either zero or non-zero 
(parallel or antiparallel to the external one). We have ruled out the
possibility of a quadrupolar splitting by checking that both the length of the RF pulse yielding
a $\pi/2$ rotation and  the recovery curve of nuclear magnetization were the same for
all lines. Moreover the observed splitting is nearly an order of magnitude larger than 
the quadrupolar one calculated on the basis of a point charge approximation.   
One notices (see Fig. 4b)
a rather sharp, but continuous, decrease of the order parameter close to $T_c$. 
An accurate determination of the critical exponent
$\beta$ would require a temperature stability better than $5\times 10^{-3}$ K which
could not be achieved with our cryogenic apparatus. Still an upper limit for $\beta$
can be estimated from our measurements which are carried out in steps of $10^{-2}$ K for $T\to T_c$.
We find that $\beta\leq 0.25$, a value compatible with a 2D Ising phase
transition, where  $\beta= 1/8$. 
It should be observed that the relative amplitude of the central and satellite lines
varies with decreasing temperature. This could be due to a  modification
of the interplanar correlation, to which $^7$Li spectrum is sensitive. To elucidate
this aspect further investigation of the collinear order is demanded. 


In conclusion, we have presented for the first time susceptibility measurements showing that 
Li$_2$VOSiO$_4$ and Li$_2$VOGeO$_4$ can be considered as prototypes of
frustrated 2DQHAF and NMR spectra demonstrating that in the former a collinear 
phase is established at
low-temperature, as predicted for $J_2/J_1 > 1/2$ \cite{Chandra2}. 
Finally $^{29}$Si NMR spectra
suggest the occurrence of a structural distortion, just above the magnetic transition, 
which lifts the degeneracy between the two collinear ground states.











\begin{figure}
\caption{a) Schematic phase diagram of a frustrated 2DQHAF on a square lattice 
as a function of the ratio
$J_2/J_1$ of the superexchange couplings. b) Structure of LSVO and LGVO projected along [001].
(Si,Ge)O$_4$ tetrahedra are in grey, VO$_5$ pyramids are in black while the grey circles indicate
Li$^+$ position. For details see Ref. 8.}
\end{figure}









\begin{figure}
\caption{Temperature dependence of the magnetic susceptibility $\chi = M/H$, for $H=0.3$
Tesla, in LSVO (closed squares) and LGVO (open circles) powders. In the inset the corresponding
temperature dependence of the derivative $d\chi /dT$ is reported.}
\end{figure}


\begin{figure}
\caption{Temperature dependence of $^7$Li NMR $1/T_1$  for $H=1.8$
Tesla along the $c$ axis in a LSVO single crystal. 
In the inset we report the corresponding Arrhenius plot of the 
experimental data for $T\leq 2.66$ K. The error bars when not reported are within
the circles. }
\end{figure}




\begin{figure}
\caption{ 
a) $^{29}$Si NMR powder spectrum in LSVO for $H=1.8$ Tesla. The dotted lines mark the position
of the peak at high and at low temperatures. b) Temperature dependence of the splitting
between the low-frequency and the high frequency peak  for $^7$Li NMR spectra in c).
The solid line indicates the behaviour expected for an order parameter with a critical exponent
$\beta=0.25$. c) $^7$Li NMR spectra for $H=1.8$
Tesla along the $c$ axis of a LSVO single crystal, in the proximity of 
$T_c=2.83\pm 0.005$ K.
}
\end{figure}










\begin{references}
\bibitem{Notachi}
The values obtained for the Van-Vleck term $\chi_{VV}$ are around 
$4\times 10^{-4}$ emu/mole
for both samples, a value typical for V$^{4+}$ in pyramidal environment, while
the Curie constant $C$ turns out to be $\simeq 0.3$ emu.K/mole for LGVO 
and $\simeq 0.33$ emu.K/mole for LSVO, 
in good quantitative agreement with what is expected for $S=1/2$.
\bibitem{NotaLi} The assumption that the lattice distortion has little affected 
the hyperfine copuling between V$^{4+}$ spins and $^7$Li nuclei is justified by
the fact that $^7$Li NMR spectrum is unaffected by the structural distortion.
\end{references}
\end{document}